\def\year{2021}\relax
\documentclass[letterpaper]{article} 
\usepackage{aaai21}  
\usepackage{times}  
\usepackage{helvet} 
\usepackage{booktabs}
\usepackage{courier}  
\usepackage[hyphens]{url}  
\usepackage{graphicx} 
\usepackage{natbib}  
\usepackage{caption} 
\usepackage[dvipsnames]{xcolor}
\usepackage[normalem]{ulem}
\nocopyright
\title{Look behind the Censorship: Reposting-User Characterization and Muted-Topic Restoration}
\author{
    Yichi Qian,\textsuperscript{\rm 1} Qiyi Shan,\textsuperscript{\rm 2} Hanjia Lyu,\textsuperscript{\rm 1}  Jiebo Luo\textsuperscript{\rm 1} \\
}
\affiliations{
\textsuperscript{\rm 1} University of Rochester\\
\textsuperscript{\rm 2} Pennsylvania State University
}

\begin{document}

\maketitle

\begin{abstract}
The emergence of social media has largely eased the way people receive information and participate in public discussions. However, in countries with strict regulations on discussions in the public space, social media is no exception. To limit the degree of dissent or inhibit the spread of ``harmful'' information, a common approach is to impose information operations such as censorship/suspension on social media. In this paper, we focus on a study of censorship on Weibo, the counterpart of Twitter in China. Specifically, we 1) create a web-scraping pipeline and collect a large dataset solely focus on the reposts from Weibo; 2) discover the characteristics of users whose reposts contain censored information, in terms of gender, device, and account type; and 3) conduct a thematic analysis by extracting and analyzing topic information. Note that although the original posts are no longer visible, we can use comments users wrote when reposting the original post to infer the topic of the original content. We find that such efforts can recover the discussions around social events that triggered massive discussions but were later muted. Further, we show the variations of inferred topics across different user groups and time frames.
\end{abstract}

\section{Introduction}
With the rapid growth of the Internet industry in China, the regulations and censorship from authorities are like the ``Sword of Damocles'' hanging over the head of Internet companies. 
The platforms that do not censor sensitive content are subject to punishment or completely removal from the Chinese market.\footnote{\url{https://supchina.com/2018/04/12/jokes-app-neihan-duanzi-shuttered-by-chinas-media-regulator-for-vulgarity/}} The most recent case is the withdrawal of LinkedIn from China because it allows politically sensitive information to be spread on users' feed~\cite{weise_mozur_2021}. 

Weibo, the largest indigenous micro-blogging service provider, is regarded as the most active social media platform that may post threat to the state autonomy over information due to its wide availability of information from individual users rather than state-run agencies~\cite{stockmann2017social}. To cope with the challenge brought by its openness, Weibo set up a complicated system to impose censorship.\footnote{\url{https://cpj.org/2016/03/the-business-of-censorship-documents-show-how-weib/}} The users on this platform invented a term ``JIA'' (In English, ``clip'') to refer to the occurrence of censorship.\footnote{\url{https://weibo.com/5668423034/J4N9BDlRy}} Such reference signifies the frequency of posts being deleted or censored, and its consequential measures such as account being muted for a certain period of time or even the closure/suspension of the account.

\begin{figure}[htbp]
  \centering
  \includegraphics[width=\linewidth]{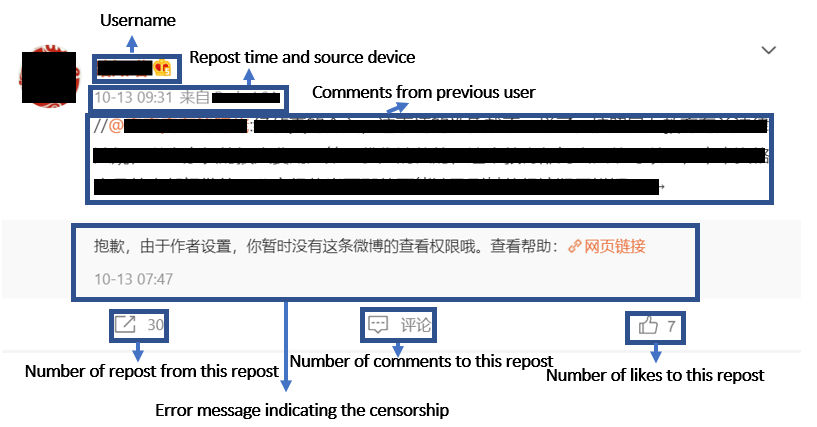}
  \caption{An example repost where the original post has been censored.}
  \label{fig: demo of single post}
\end{figure}

In this paper, we focus on (1) depicting the user characteristics of those who repost the posts that are later deemed inappropriate and censored, and (2) inferring and analyzing the topics of these censored posts. To achieve both goals, we first create a pipeline to collect user IDs, their corresponding profile information, as well as their reposts. These data are all publicly available. Next, we characterize the users who repost the posts that are later censored. Although the original posts have been censored, the comments users wrote when reposting the original posts (such as the ``Comments from previous user'' in Figure~\ref{fig: demo of single post}) are still available. By applying the Chinese word segmentation tool Jieba~\cite{sun2012jieba} and Latent Dirchlet Allocation (LDA)~\cite{blei2003latent} to the comments, we infer the topics of the original posts. We obtain a better understanding of the relation between social media users and information operations such as censorship. To summarize, our contributions are as follows:

\begin{itemize}

    \item We discover several important insights regarding the users who repost censored contents, including: 1) Male users are more likely to repost contents that are later censored while female users are more active in reposting; 2) Reposts from web browsers are twice likely to contain censored posts; 3) Social capital (e.g., number of followers and the verification status) is negatively correlated with the likelihood of reposting posts that are later censored.
    \item We infer the topics of the posts that are censored by analyzing the still-available user comments which were written in the reposts. We show the focus varies across different user groups and time frames based on the topic inference.
    
    
\end{itemize}

\section{Related Work}

Researchers have studied the censorship in Chinese social media from different perspectives. \citet{king2013censorship} conducted a comprehensive study  on mainstream blog websites and bbs sites. They discovered that the goal of censorship is to cut social ties and thus prevent collective actions. \citet{liu2021more} found that multimedia posts suffer  from ore scrutiny than plain-text posts. \citet{arefi2019assessing} implemented a multi-model approach, combining computer vision and NLP techniques, to predict whether a post would be censored. They showed that political posts are more likely to be censored. Additionally, \citet{ng2018detecting} used linguistic features and Naive Bayes classifier to predict the likelihood of censorship. \citet{kenney2021user} used pure user demographic information to build a binomial regression model to predict the degree of censorship with the datasets collected in 2012. Although they abandoned their model due to poor performance, they inferred that male verified users were very likely to post offensive content. 


Previous work around content mining on Weibo~\cite{bao2013popularity, wang2020concerns} either has a limited scope on topics or the users groups or both, while this study does not limit the scope of our crawler from the beginning. In addition, to our best knowledge, our study presents the first attempt of topic inference/restoration from comments on reposts, while most linguistic feature approaches in the censored topic extractions are based on the censored content itself, which would need a second-time query. 

\section{Data Collection}

Weibo supports massive amount of daily traffic, and restricts web crawlers from scraping information freely. To overcome this obstacle, we design a framework that allows us to collect a relatively representative Weibo dataset. The data collection procedure is composed of (1) collecting user IDs, and (2) collecting user characteristics and posts.

\subsection{Collecting User IDs}

\begin{figure}[h]
  \centering
  \includegraphics[width=0.9\linewidth]{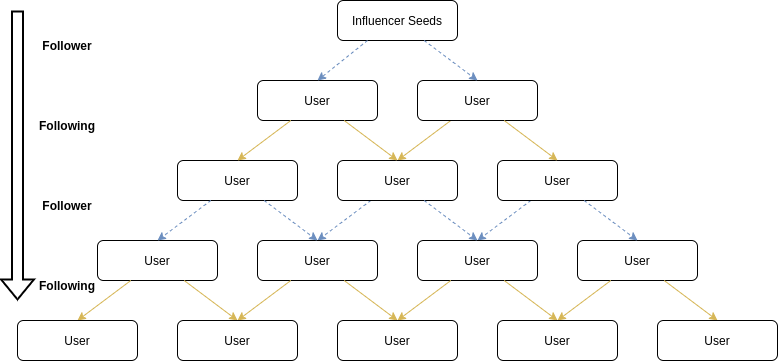}
  \caption{Diagram of searching for users.}
  \label{fig:user_id_scraping_order}
\end{figure}

First, we create a seed user set. Specifically, when we create a new Weibo account, Weibo will randomly recommend 80 influencers to us so that we can decide whether we follow them or not. These 80 influencers are related to almost all major fields, including ``interesting video bloggers'', ``Weibo hot-topic'', celebrities, anime, electronics, sports, general science, humor, fashion, politics, sports, education, vehicles, gourmets, finance, pets, fashion, culture, and lifestyle. We use these 80 influencers to construct the seed user set. Next, we search for users by a shifting mechanism which is shown in Figure~\ref{fig:user_id_scraping_order}. We crawl the followers of these 80 influencers, denoted by $User\_Set_{1}$. We then crawl the users that are followed by users in $User\_Set_{1}$, denoted by $User\_Set_{2}$. After obtaining $User\_Set_{2}$, we crawl the followers of the users in $User\_Set_{2}$, denoted by $User\_Set_{3}$. We keep crawling users among the follower list and the following list until there are no more new users. One disadvantage of only crawling the users from the following list is that the Weibo users with a larger social capital (e.g., more followers) tend to follow each other more frequently, which may create sampling bias. However, only crawling the users from the follower list may create other drawbacks. For example, due to the high ratio of Weibo ghost accounts being used to create fake followers~\cite{zhang2016discover}, the dataset that is constructed only based on the users from the follower list may contain a high proportion of ghost accounts. To account for the trade-off between using only the follower list or the following list, we choose to crawl users from both of them. In the end, we collect 1,028,264 unique Weibo user IDs. 

To confirm that our seeding method does not introduce undue selection biases to the collected dataset, we further seed another 80 celebrities and collect a second smaller dataset with 50,000 users for robustness verification. We conduct the same experiments using the first and second datasets. The findings do not fundamentally change. 

\begin{figure}
    \centering
    \includegraphics[width=\linewidth]{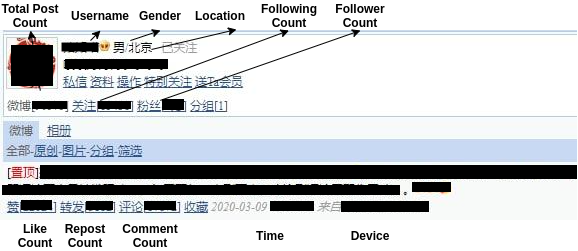}
    \caption{Available User Information.}
    \label{fig:user_information_showcase}
\end{figure}

\subsection{Collecting User Characteristics and Posts}
In Weibo, user ID is a unique identifier for each user. With user IDs, we are able to access users' publicly available characteristics and posts by visiting their home page URL through \emph{https://weibo.cn/u/[user ID]}. From the homepage of each user (see Figure~\ref{fig:user_information_showcase}), we crawl the user characteristics including username, gender, location, post count, follower count, and following count. All the publicly available posts and the corresponding attributes are collected including the post content, published time, device, ``like'' count, repost count, and comment count. Around 20 million posts are retrieved. 

To determine whether a post has been censored by the operator/authorities when we encounter error messages in retrieving the original post, we take a close look at all the error messages Weibo issues. If the error message is in one of these circumstances, it means the post was censored forcibly:
\begin{enumerate}
    \item The error message indicates that the account was reported for violating community rules;
    \item The error message indicates that the the non-visibility is caused by the permission set by the poster;
    \item The error message indicates that the post has been reported by multiple users for violating community rules;
    \item The error message contains a link prompting user to look for further guidance.
\end{enumerate}
The second seems to relate the non-visibility to user's subjective choice, while in fact it is not. If a user takes the initiative to set the visibility, then the error message would clearly clarify to which visibility level the user set. For example, if the user set the visibility to ``6 months'', then the error message would mention that the post is set as ``only visible for the first 6 months''.

\section{Descriptive Analysis}


Table~\ref{tab:summary} provides a general summary of the dataset we have collected. Among all the users, we find that around 450,000 users have not reposted any posts. There are multiple reasons behind this: some of them may be ghost accounts~\cite{zhang2016discover} and these accounts are often very silent; it could also be that a user solely uses Weibo to record personal feelings so rarely does this user repost; another explanation is that the user's account has been muted due to inappropriate posts. 
Once a user is muted, not only can he/she not send any post, but all his/her past posts/reposts are also set hidden from other users. Further, among the users with at least one repost, we find 179,307 (31.21\%) of them have reposted at least one post that is later censored.
\begin{table}[h]
\tiny
\centering
  \caption{Summary of the collected data.}
  \label{tab:summary}
  \begin{tabular}{c c l}
    \toprule
    Data &Count\\
    \midrule
    Reposts & 19,428,137 \\
    Reposts with censored original posts & 553,509 \\
    Users & 1,028,264\\
    Users with at least one repost & 574,356 \\
    Users whose reposts contain censored content & 179,307\\
  \bottomrule
\end{tabular}
\end{table}


In terms of reposts, the reposts with censored original content account for roughly 2.8\% of the total reposts, which is \textit{consistent} with the findings of \citet{ng2020linguistic}, who used a combination of third-party data and self-collected data, indicating a good performance of our data collection framework.

Because users have different number of reposts, the absolute number of reposts with censored original content alone is not a good measure to analyze how frequently a user reposts content that is later censored. Hence, we use the \textit{ratio of reposts with censored content} by normalizing the number of reposts with censored original post by the number of reposts. The calculations are as follows \[ \frac{\# Reposts\ with\ Censored\ Content}{\#  Reposts }\]

Our study focuses on the users with at least one repost. For convenience, we refer to this group of users as active users and the study from this point will mainly focus on this group of users.

\subsection{Gender}

The majority of active users on Weibo are women (66.5\%). This finding disagrees with the gender distribution (Men account for 57\% of the total users) found by other researchers~\cite{seema_2021}. However, the reason behind this huge gap might be because that male users repost less. The average \textit{ratios of reposts with censored content} of men and women is 0.04 and 0.02, respectively. Men are more likely to repost posts that are later censored. This finding is \textit{consistent} with the conclusion of \citet{kenney2021user}.




\subsection{Social Capital}

\subsubsection{Verification Status}

The verification of an account on Weibo is very much like Twitter, but with a much lenient criteria~\cite{seo_2019}. The major difference is that Twitter focuses more on the contributions a user have for certain area(s) and whether the user has continuously make such contributions, while Weibo only requires that the user needs to be followed by two or more verified users and the numbers of followings and followers need to exceed 50. 
In our dataset, the ratio of verified accounts versus unverified accounts is roughly 1:13. In terms of the comparison of \textit{ratios of reposts with censored content}, 
the users from the unverified group tend to have more reposts with censored content (0.03), compared to the average ratio of verified accounts (0.02). The stricter punishment against verified users,\footnote{\url{https://www.economist.com/analects/2013/09/16/humiliating-the-big-vs}} especially those with many followers, explains such gap. For users that have accumulated millions of followers over years, the cost of the suspension is expensive. 



\subsubsection{Influence Power}
To further examine how social capital affects user's repost pattern, we introduce a new metric -- ``Influence Power'' on a scale from 1 to 10 degrees. Each degree means 10\% quantile. For example, if a user falls into the second degree group, then it means that his/her number of followers fall within the range of 10\% quantile and 20\% quantile among the distribution of number of followers for all active users. In our dataset, the users of the 8th degree group have the most ``Influence Power''.
The range of number of followers in the 8th degree group is between 196 to 367. Therefore, the spike at the 8th degree group suggests that users with around 200 followers tend to repost content that is later censored. 




\subsection{User Device}
Weibo records the detailed information of the devices that the users use to repost/post. We aggregate the devices to the brand level, and investigate whether or not it has any impact on the ratio of reposts with censored content. 
\emph{Apple} monopolizes as the top brand from which reposts are made, surpassing the summation of the next three brands (\emph{Android}, \emph{Honor}, \emph{Huawei}). Interestingly, from the 2021Q2 market report~\cite{team_2021}, \emph{Apple} ranked the fourth place in shipments, while \emph{Vivo}, \emph{Oppo}, \emph{Xiaomi} ranked from the first to the third, respectively. 
The reason of the difference between \emph{Apple}'s market share and the Weibo usage can be that users of both \emph{Apple} and Weibo overlap heavily, while users using other devices do not use Weibo as much. With respect to the average \textit{ratio of reposts with censored content} by different devices. \emph{Web} has the highest ratio which indicates that reposts from Web browsers are more likely to contain content that is censored. Among the remaining brands, \emph{Xiaomi} takes the lead, but with little difference with its runner-ups. 




\section{Feature Importance Analysis}
\subsection{Model Selection}
In this section, we intend to shed light on the influence of the aforementioned features on whether posting a repost contains censored content (for simplicity, we will refer it as ``censored repost'' in the following part of this paper). To measure the influence, we construct a decision tree-based model and analyze the feature importance. 
For the tree model, we pick XGBoost~\cite{chen2016xgboost} over other boosting models because it outperforms other tree models in the experiment on heterogeneous data by \citet{gorishniy2021revisiting}.

\subsection{Training}
To feed our data into the model, we first one-hot encode our categorical variables. The features of interests are: device, user gender, number of posts, number of followers, number of followings, verification status, province and province GDP. 

The dataset itself is extremely unbalanced, with only 2.8\% of the reposts containing censored post. To construct a relatively balanced training set, we randomly sample 400,000 censored repost and 800,000 uncensored reposts. The number of uncensored posts is set to 800,000 in an attempt to capture as much information as possible while alleviating the data imbalance issue. 
For the testing set, we use the remaining 153,509 censored reposts and randomly sample 300,000 uncensored reposts. We set the learning rate at 0.02. We achieve an AUC score of 0.639 on the testing dataset. 

\begin{figure}[h]
    \centering
    \includegraphics[width=0.8\linewidth]{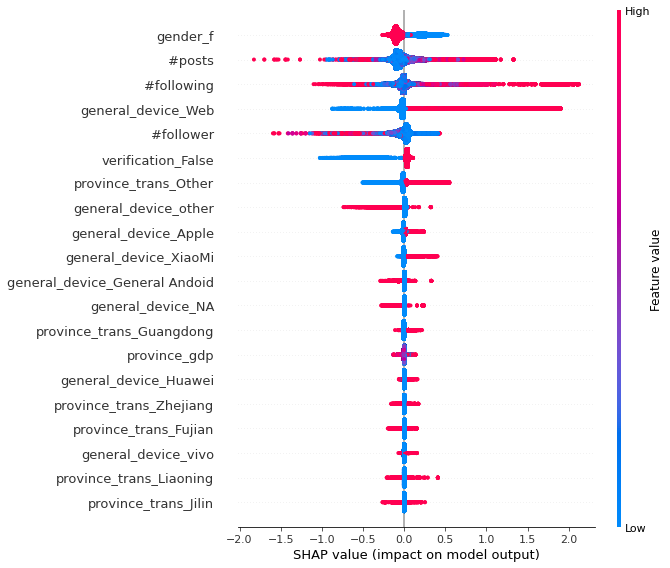}
    \caption{SHAP approach of summarizing effects of features (we translate provinces from Chinese to English, so there is a ``trans'' in province variables; general\_device\_NA means the device information is hidden by users).}
    \label{fig:shap summary}
\end{figure}


\subsection{Feature Interpretation}
To better interpret the model, we use SHAP~\cite{lundberg2020local2global}. The SHAP values are shown in Figure~\ref{fig:shap summary}. The most important feature is ``gender\_f'', and we can see that high value of this feature imposes a negative impact on the output, and vice versa. This verifies our finding in the previous section, that men are more likely to post censored reposts. For the number of posts and the number of followings, they follow a similar pattern: two sides are most filled up by higher values and the middle is crowded by lower values, indicating that higher values can affect the output both negatively and positively while lower values have neutral effect. As for the usage of the device, we find that reposts from web browser are more likely to contain censored contents, which confirms the aforementioned discussions. The distribution of the SHAP values of the number of followers and  ``verification\_False'' shows that as the value gets higher, the more negative effect it has on the output variable. A higher number of followers or being a verified user is negatively related to the chance of reposting censored content. This finding confirms our observation from the previous section as well. 

\section{Topic Inference}
To find out what content are more likely to be censored, we use LDA~\cite{blei2003latent} to extract the topics. To measure the quality of an LDA, we use the coherence score ($C\_v$) and choose the number of topics with the highest score. Table~\ref{tab:general_censored} shows the top keywords per topic. Weibo users mostly talk about the stars of the entertainment industry such as ``Wu Yifan'', ``Xiao Zhan'', ``Zhang Zhehan'', etc. For example, ``Wu Yifan''(also known as Kris Wu), a top pop star, has triggered hot discussions after he was accused of doping and sexual assault by a victim. Later, after the local police arrested and detained him,\footnote{\url{https://www.globaltimes.cn/page/202108/1231619.shtml}} Weibo started to censor everything that was related to him, as well as the closure of his account. The appearance of the entertainment-related discussions can be attributed to the recent crackdown on celebrities whose behavior exceeded authorities' bottom-line.\footnote{\url{https://www.scmp.com/lifestyle/entertainment/article/3149435/zhao-wei-kris-wu-zhang-zhehan-chinese-stars-hit-chinas}} In the past, these celebrities' posts were often reposted frequently by their fans. After the closure/suspension of the celebrities' accounts, these reposts then become the one without visible original contents. Meanwhile, the flooding in Henan\footnote{\url{https://www.reuters.com/world/china/death-toll-flooding-chinas-henan-province-rises-302-2021-08-02/}} also involves many discussions, many of which are about seeking assistance and reporting situation. 

\begin{table*}[!htbp]
\tiny
  \caption{LDA results of all comments on censored posts.}
  \label{tab:general_censored}
  \begin{tabular}{l p{16cm}}
      \hline
      Topic&Words per topic\\
      \hline
		1& say (0.07), do (0.04), expand (0.02), fan (0.02), Wang Yibo (0.01), scold (0.01), think (0.01), search (0.01), feel (0.01), find (0.01)\\ 
		2& stance (0.06), vote for (0.06), options (0.05), Wu Yifan (0.04), participate (0.03), initiate (0.02), vote (0.02), studio (0.02), ya (0.02), happiness (0.02)\\
		3& Henan (0.12), rainstorm (0.12), mutual assistance (0.09), trapped (0.02), elder (0.02), Zhengzhou (0.01), ask for help (0.01), urgent need (0.01), Xinxiang (0.01), lost contact (0.01) \\
		4& picture (0.04), check (0.03), video (0.02), elder brother (0.02), please (0.01), Xiao Zhan (0.01), help (0.01), world (0.01), money (0.01), studio (0.01)\\ 
		5& forward (0.05), think (0.02), hope (0.02), focus on (0.02), deliver (0.02), smoke (0.02), era (0.02), Yan Haoxiang (0.01), case (0.01), grateful (0.01)\\ 
		6& Bo (0.05), micro (0.04), change (0.04), lottery (0.02), platform (0.02), grass (0.02), link (0.01), fucking (0.01), forward (0.01), pregnant woman (0.01) \\
		7& comment (0.04), die (0.04), smile (0.04), love (0.04), fan (0.04), eat (0.03), with pictures (0.02), come on (0.02), Hunan satellite TV (0.01), picture (0.01) \\
		8& ah ah ah (0.05), like (0.04), thanks (0.04), mom (0.02), new (0.02), teacher (0.01), cry (0.01), shoot (0.01), character (0.01), great (0.01) \\
		9& cute (0.03), china (0.03), support (0.02), Zhang Zhehan (0.02), child (0.01), apologize (0.01), woman (0.01), nation (0.01), gong jun (0.01), stand (0.01) \\
		10& Wu Yifan (0.13), Mr (0.05), endorsement (0.03), spokesperson (0.03), road (0.02), expect (0.02), easy Vuitton (0.02), brand (0.02), force (0.02), worldwide (0.02) \\
      \hline
\end{tabular}
\end{table*}

\subsection{Topics of the Subgroups}
In previous sections, we have found that gender, social capital and the device are related to whether or not a Weibo user reposts censored content. To gain more insights, we conduct further topic analysis with regarding to these factors. More specifically, we apply the LDA to the censored reposts by female users, unverified users, and web browser users, respectively.

\subsubsection{Female Users}
The topics of censored reposts by female users are more aggregated. Seven topics yield the highest coherence score. Except for the topics regarding flooding and entertainment stars, female users also talk about women, laws, etc., which are not found in the topics extracted from the general corpus.

\subsubsection{Unverified Users}
The topics extracted from the censored reposts by unverified users are very much the same as the topics extracted from the general corpus, except that in one topic, it shows many Samaritan words, such as ``engage in', ``find'', ``help'', ``expand (means spreading useful information in Chinese)'', etc., which shows that unverified users are more willing to play an active role on the platform.

\subsubsection{Web Browser Users}
Three major differences are observed in the topics of the censored reposts by web browser users: (1) web browser users attend online lottery frequently; (2) topics also include shopping of top-trended items on ``T-Mall'' - an online shopping website resembles Amazon; (3) these reposts show much participation into online surveys.

\section{Temporal analysis}

We aggregate the ratios to a monthly average and observe the points of changes. 
There are two major spikes, one is in the end of 2019 and the other is in September 2020. 


To figure out what happened in late 2019, we select comments sent from November 2019 to February 2020 and apply the LDA model. Most topics are related to COVID-19. Apart from these pandemic-related topics, we have also observed the existence of ``Wu Yifan'', suggesting that ``Wu Yifan'' has long been a frequent topic.

We examine the comments from September 2020 to December 2020 in the same way. After almost a year since the outbreak of pandemic, the COVID-19 related topics do not appear in the LDA results of the cencored reposts. Rather, there are many topics that are about ``Wu Yifan''. To be specific, the topics regarding him show his social activities, including car racing and commercial endorsements. After searching reports about his activities during that period, we find that he was racing cars for Porsche,\footnote{\url{https://planetporsche.org/2020/10/30/porsche-sports-cup-returns-to-shanghai-for-momentous-porsche-sportscar-together-day/}} and attending fashion shows as ambassador of Louis Vuitton. \footnote{\url{https://www.dazeddigital.com/fashion/article/51091/1/kris-wu-autumnwinter-2020-issue-virgil-abloh-louis-vuitton}} Apart from ``Wu Yifan'', we also observe topics related to ``live stream'', which is also a new topic that has not been observed in previous analysis.

Lastly, we study the comments in July 2021, when the scandal of ``Wu Yifan'' broke out, and the flooding in Henan province took place. We see many topics regarding the flooding, at least three of which are reposting help-seeking and status-update message. In terms of topics surrounding``Wu Yifan'', there is one expressing concerns and attention on the progress of the matter, one talking about laws and expressing astonishment and the others are just admiring comments from the fans. 

\section{Conclusion}
This paper presents an improved web crawling pipeline to study user characteristics behind the repost with censored posts and the related topics inferred from still available comments related to now invisible censored contents. The study is conducted at a large scale and covers reposts in a span of over two and a half years. 

In terms of user characterization, we show that 1) female users have a higher tendency of reposting while male users are more bold in reposting sensitive contents;  2) social capital acts as a heavy leverage when users repost contents, and as social capital increases, user's vigilance in reposting also increases; 3) reposts from web browsers have almost the twice likelihood than other devices in including censored original posts. Based on topic inference, we show the focus varies across different user groups and time frames.

Our study has several implications for the study of social media in countries with censorship policies. First, we propose a method to maximally restore the topics from comments even though the original posts are unavailable. This approach can save much time for researchers if they are concerned about topic extraction from missing data due to censorship. Second, we show the user portraits with regards to the likelihood of a user reposting content that are later censored from a large-scale dataset, which gives the broader society a sense of who are the users that are more actively involved with the discussion of sensitive topics.

By no means is this paper a comprehensive study on the censorship on Weibo. There are many other types of censorship that we do not cover. For example, \citet{liu2021more} mentioned the multimedia censorship including images and videos. Recently Weibo has implemented a new mild way of censoring posts by converting the visibility of a Weibo post without deleting the whole post.\footnote{https://kefu.weibo.com/faqdetail?id=21092} In addition, this study focuses on the reposts of which the original content is already censored or muted because censorship/information blockage is the focal point of this paper. In the next phase of our work, we plan to delve deeper into the broader corpus that we have collected.

Finally, this paper does not make any judgement on the policies each country/platform implemented in managing their social media. The purpose of this research is purely to study the characteristics of a certain group of users and look for methods of topic extraction under difficult circumstances.

\section{Broader Impact and Ethical Consideration}
Although not originally intended, this work reveals a way to restore the muted topics that are restrained from public discussion and dissemination, which may lead to a fortified action from the authorities to censor the flow of information. During the crawling process, we only crawl publicly available data on Weibo. In addition, we anonymize all users we have collected during the research so that users are not identifiable since this study does not examine any specific user. 

\bibliography{sample-base}
\bibstyle{aaai}
\appendix

\end{document}